\documentclass[conference]{IEEEtran}
\IEEEoverridecommandlockouts
\usepackage{cite}
\usepackage{amsmath,amssymb,amsfonts}
\usepackage{algorithm}
\usepackage{algorithmic}
\usepackage{graphicx}
\usepackage{textcomp}
\usepackage{xcolor}
\usepackage{bm}
\newcommand\myciteup[1]{{\setcitestyle{square,super}\cite{b1}}}
\def\BibTeX{{\rm B\kern-.05em{\sc i\kern-.025em b}\kern-.08em
    T\kern-.1667em\lower.7ex\hbox{E}\kern-.125emX}}
\begin{document}

\title{Joint Millimeter Wave and Microwave Wave Resource Allocation Design for Dual-Mode Base Stations\\
}
\author{\IEEEauthorblockN{Biqian Feng, 
		Zhijun Liao, Yongpeng Wu, Juening Jin, Derrick Wing Kwan Ng, Xiang-Gen Xia, and Xinbao Gong}
	
\thanks{The work of Y. Wu is supported in part by  the National Science Foundation (NSFC) under Grant 61701301 and Young Elite Scientist Sponsorship Program by CAST.}
\thanks{B. Feng, Y. Wu, J. Jin, and X. Gong are with the Department of Electronic Engineering, Shanghai Jiao Tong University,
	Minhang 200240, China (e-mail: fengbiqian@sjtu.edu.cn; yongpeng.wu@sjtu.edu.cn; jueningjin@gmail.com; xbgong@sjtu.edu.cn)(Corresponding author: Yongpeng Wu.).}

\thanks{Z. Liao is with Technology and ICT Department, Jiangxi Electric Power Company, No.666 Hubin East Road, Qingshan Lake District, Nanchang, Jiangxi (Email: l: ycglzj@163.com)}

\thanks{D. W. K. Ng is with the School of Electrical Engineering and
	Telecommunications, University of New South Wales, Sydney, N.S.W.,
	Australia (E-mail: w.k.ng@unsw.edu.au).}

\thanks{X.-G. Xia is with the Department of Electrical and Computer Engineering, University of Delaware, Newark,
	DE 19716, U.S.A. (e-mail: xxia@ee.udel.edu).}
	
}

\maketitle

\begin{abstract}
In this paper, we consider the design of joint resource blocks (RBs) and power allocation for dual-mode base stations operating over millimeter wave (mmW) band and microwave ($\mu$W) band. The resource allocation design aims to minimize the system energy consumption while taking into account the channel state information, maximum delay, load, and different types of user applications (UAs). To facilitate the design, we first propose a group-based algorithm to assign UAs to multiple groups. Within each group, low-power UAs, which often appear in short distance and experience less obstacles, are inclined to be served over mmW band. The allocation problem over mmW band can be solved by a greedy algorithm. Over $\mu$W band, we propose an estimation-optimal-descent algorithm. The rate of each UA at all RBs is estimated to initialize the allocation. Then, we keep altering RB's ownership until any altering makes power increases. Simulation results show that our proposed algorithm offers an excellent tradeoff between low energy consumption and fair transmission.

\end{abstract}


\section{Introduction}
The tremendous growth of wireless devices stimulates the development of the fifth-generation (5G) communication network. 5G network is expected to support a connection density up to $10^6/km^2$ \cite{b1}, which is about 10 times higher than that of the fourth-generation (4G) network. In order to support the tremendously increased throughput requirement, 5G wireless network makes use of both microwave ($\mu$W) band and millimeter wave (mmW) band. Therefore, resource blocks (RBs) allocation over these two bands have become a fundamentally important topic.

The conventional scheduling algorithms include round robin (RR), maximum carrier to interference ratio (MAX C/I), and proportional fair (PF) \cite{b2}. Besides, a number of novel scheduling algorithms have been proposed, such as context-aware algorithm \cite{b3}, multiuser adaptive orthogonal frequency division multiplexing (MAO) scheme \cite{b4}, and weighted sum power minimization (WSPmin) \cite{b5}. In particular, RR offers fairness among user applications (UAs) in radio resource assignment, but it degrades the whole system throughput considerably. Besides, MAX C/I takes full account of system throughput, but ignores the resource allocating fairness. PF makes a tradeoff between fairness and system throughput, but it does not involve the power allocation procedure. Context-aware algorithm considers that UAs with bad CSI wastes network resources and experiences a transmission delay. MAO only considers power allocation in one slot. WSPmin can only be used in orthogonal frequency division multiple access (OFDMA) efficiently \cite{b5}. With this in mind, we hope to design a power allocation algorithm which can achieve a good tradeoff between fairness and system performance in multiple time slots.

The main contribution of this paper is to take fairness and power allocation into consideration. Besides, the operating of allocation system is decomposed into three stage, including grouping, allocating in mmW, and allocating in $\mu$W. More specifically, group-based (GB) algorithm is proposed to divide UAs into multiple time slots and reduce the competition among similar UAs sharing the same expectation for RBs. Then, we apply estimation-optimal-descent (EOD) algorithm to reduce power consumption in each group. GB algorithm and EOD algorithm cooperate with each other to complete power allocation effectively.

The rest of this paper is organized as follows. In Section \uppercase\expandafter{\romannumeral2}, system model and problem formulation are introduced. In Section \uppercase\expandafter{\romannumeral3}, we propose GB algorithm to assign UAs to multiple slots. Section \uppercase\expandafter{\romannumeral4} provides EOD algorithm to allocate RBs over $\mu$W and mmW in detail. Simulation results and discussion are given in Section \uppercase\expandafter{\romannumeral5}. Section \uppercase\expandafter{\romannumeral6} concludes the paper.

\section{System Model And Problem Formulation}
We consider a multiuser downlink dual-mode transmission system, where BS simultaneously operates over both $\mu$W and mmW bands. Meanwhile, user equipments (UEs) are equipped with interfaces which can receive information from both frequency bands. Transceivers equipped with large antenna arrays achieve an overall beamforming gain to overcome path loss over mmW band \cite{b6}\cite{a1}\cite{a2}\cite{a3}\cite{a4}. Under this model, BS is located at the center of the cell and $M$ UEs are deployed randomly within a circular cell with radius $d$. It is assumed that UE $m$, located at $(d_{m},\theta_{m})$, runs $\kappa_{m}$ UAs. So, there are $\sum_{m=1}^M\kappa_{m}$ UAs needed to be served successfully.
\subsection{Channel Model}
Over $\mu$W band, we adopt OFDMA scheme. The duration time for each slot is $\tau$ in downlink transmission. We assume that there are $K_1$ RBs available to be allocated to all UAs. Then, the rate of UA $n$ at RB $k$ and time slot $t$ is given by
\begin{equation*}
R_{nkt}^{(1)}=\omega_{1}\log_{2}\left(1+\frac{p_{1nkt}\left|g_{kt}\right|^{2}10^{-0.1L_{1}(d_{n})}}{\omega_{1}N_{0}}\right),
\end{equation*}
where $\omega_{1}$ denotes the bandwidth of each RB at $\mu$W band; $p_{1nkt}$ is the power allocated to UA $n$; $g_{kt}$ represents the Rayleigh fading channel coefficient; $L_{1}(d_{n})$ denotes the large-scale path loss over $\mu$W band.

For mmW, we adopt time division multiple access strategy due to its spectrum/bandwidth flexibility and low cost for small cells \cite{b7}. For TDMA scheme when the number of UAs is fixed, the transmission time $\tau^{'}$ is determined accordingly. Then, the rate of UA $n$ at RB $k$ and time slot $t$ is given by
\begin{equation*}
R_{nkt}^{(2)}=\omega_{2}\log_{2}\left(1+\frac{p_{2nkt}\psi(d_{n})\left| h_{kt}\right|^{2}10^{-0.1L_{2}(d_{n})}}{\omega_{2}N_{0}}\right),
\end{equation*}
where $\omega_{2}$ denotes the bandwidth of each RB at mmW band; $\psi(d_{n})$ represents the beamforming gain that UA $n$ achieves over mmW band; $p_{2nkt}$ is the power allocated to UA $n$; $h_{kt}$ denotes the Rician fading channel coefficient; $L_{2}(d_{n})$ denotes the large-scale path loss over mmW band.

As long as UA $n$ is assigned, it can obtain the entire $K_{2}$ RBs in the mmW. Therefore, the rate of UA $n$ at time slot $t$ is given by
\begin{equation*}
\begin{aligned}
R_{nt}^{(2)}&=\sum_{k=1}^{K_{2}}R_{nkt}^{(2)}\\
&=\sum_{k=1}^{K_{2}}\omega_{2}\log_{2}\left(1+\frac{p_{2nkt}\psi(d_{n})\left| h_{kt}\right|^{2}10^{-0.1L_{2}(d_{n})}}{\omega_{2}N_{0}}\right).
\end{aligned}
\end{equation*}

The model of large-scale path loss has the following form \cite{b8}:
\begin{equation*}
PL[dB](d) = \alpha+10\beta\log_{10}(d)+X_{\sigma},
\end{equation*}
where $d$ is the distance in meters, $\alpha$ and $\beta$, related to frequency and distance, are determined with a least squares fit to the measured data. $X_{\sigma}$ is the shadow fading term.

\subsection{QoS and Groups}
\newtheorem{Definition}{Definition}
\begin{Definition}
	The quality-of-service (QoS) class $\mathcal{Q}_{T}$ means a set of UAs that can tolerate maximum $T$ time slots. They will experience outage when the transmission is over $T$ slot.  \cite{b9}.
\end{Definition}

\begin{Definition}
	The Group $\mathcal{G}_{Tt}$ is defined as a set of UAs in the QoS class $\mathcal{Q}_{T}$ that are assigned to the time slot $t$ under a certain demand. UAs in the same group are served simultaneously.
\end{Definition}

Due to the constraint of time and RBs, all UAs cannot be served simultaneously. Assume that the total number of QoS class is $P$ and each UA must be allocated to a certain QoS class and group, i.e., $\sum_{m=1}^M\kappa_m = \sum_{T=1}^P\left|\mathcal{Q}_T\right|$ and $\mathcal{Q}_T = \bigcup_{t=1}^T\mathcal{G}_{Tt}$. Each UA can only be scheduled at one slot, i.e., $\mathcal{Q}_{T^{(1)}} \cap \mathcal{Q}_{T^{(2)}}=\emptyset$ and $\mathcal{G}_{Tt^{(1)}} \cap \mathcal{G}_{Tt^{(2)}}=\emptyset$.

In scheduling decision, we will take $a_{nkt}$ and $S_{nt}$ to determine the allocation over both bands. $a_{nkt}$ is the binary indicator of allocation in $\mu$W. $a_{nkt}=1$ if RB $k$ of $\mu$W is allocated to UA $n$ at time slot $t$, otherwise $a_{nkt}=0$. $S_{nt}$ is the binary indicator of allocation  in mmW. Similarly, $S_{nt}=1$ if UA $n$ is allocated to mmW at time slot $t$, otherwise $S_{nt}=0$.
\subsection{Problem Formulation}
Our goal is to design an effective scheme to minimize power for the whole QoS class $\mathcal{Q}_{T}$ in low SNR situation. Some notations are shown in TABLE I. The scheduling problem is formulated as:
\begin{subequations}
\begin{align}
&\mathop{\min}_{a_{nkt},R_{nkt}^{(1)},S_{nt},R_{nkt}^{(2)}} \sum_{t=1}^{T}\sum_{n=1}^{N}\left[\sum_{k=1}^{K1}\left(a_{nkt}p_{1nkt}\right)+S_{nt}\sum_{k=1}^{K2}p_{2nkt}\right]\nonumber\\
&\text{s.t.} \quad\sum\limits_{n=1}^{N} a_{nkt}\leq 1,\quad\forall k\in\mathcal{K}_1,\forall t\in\mathcal{T},\\
&\quad\quad\sum\limits_{n=1}^{N} S_{nt} = N^{'},\quad\forall t\in\mathcal{T},\\
&\quad\quad\tau\sum\limits_{k=1}^{K_1}{a_{nkt}R_{nkt}^{(1)}}\geq(1-S_{nt})b_{nt}^{req},\forall n\in \mathcal{Q}_{T},\forall t\in\mathcal{T},\\
&\quad\quad\tau^{'}\sum\limits_{k=1}^{K_2}R_{nkt}^{(2)}\geq S_{nt}b_{nt}^{req},\forall n\in \mathcal{Q}_{T},\forall t\in\mathcal{T},\\
&\quad\quad \mathcal{A} = \{a_{nkt}|a_{nkt}\in\{0,1\},\forall n\in \mathcal{Q}_{T},\forall k\in\mathcal{K}_1,\forall t\in\mathcal{T}\},\\
&\quad\quad \mathcal{S} = \{S_{nt}|S_{nt}\in\{0,1\},\forall n\in \mathcal{Q}_{T},\forall t\in\mathcal{T}\},\\
&\quad\quad a_{nkt}S_{nt}=0,\quad\forall n\in \mathcal{Q}_{T},\forall k\in\mathcal{K}_1,\forall t\in\mathcal{T},\\
&\quad\quad\sum\limits_{t=1}^{T} \left|\mathcal{G}_{Tt}\right| = N.
\end{align}
\end{subequations}
 
\begin{figure}[htbp]
	\centerline{\includegraphics[scale=0.2]{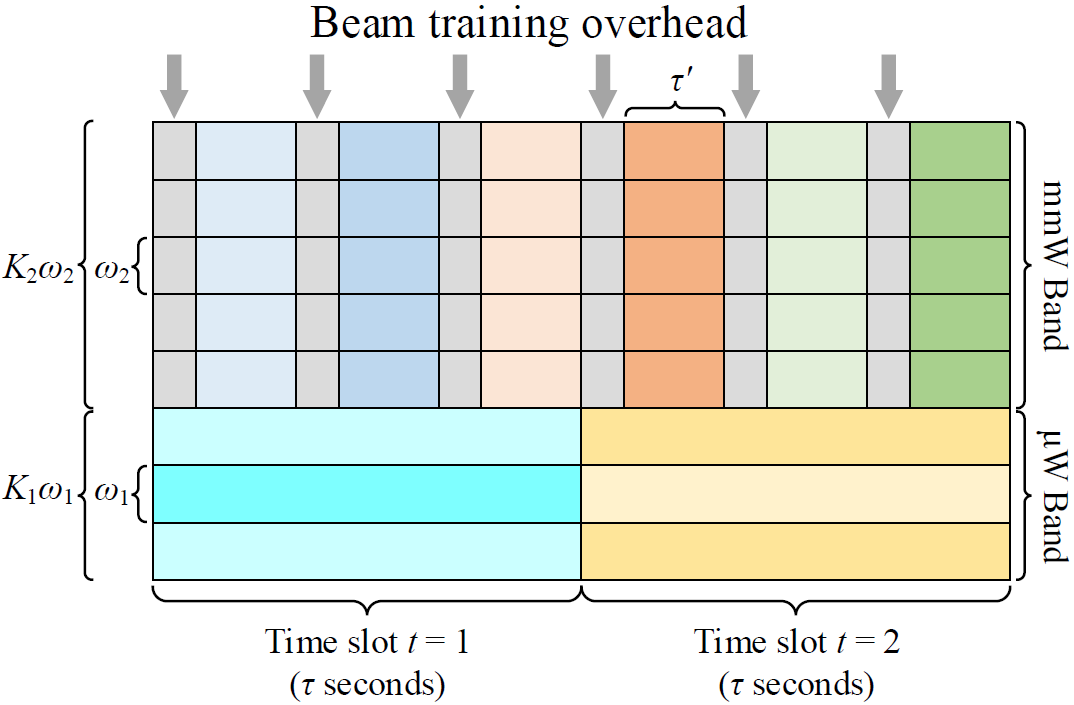}}
	\caption{Different colors represent different UAs}
	\label{fig:2}
\end{figure}

The overall allocation scheme over both frequency bands is shown in Fig.~\ref{fig:2}.

\begin{table}
	\centering
	\caption{Notations}
	\begin{tabular}{ |c|c| }
		\hline  
		Notations  & Description \\   
		\cline{1-2}
		$K_1$      & Numbers of frequency band in $\mu$W \\   
		\cline{1-2}
		$\mathcal{K}_1$ & Sets of frequency band in $\mu$W \\
		\cline{1-2}
		$K_2$      & Numbers of frequency band in mmW \\  
		\cline{1-2}
		$\mathcal{K}_2$ & Sets of frequency band in mmW \\ 
		\cline{1-2}
		$N$      & Numbers of UAs needed to be allocated \\ 
		\cline{1-2}
		$N^{'}$      & Numbers of UAs allocated in mmW at each slot\\ 
		\cline{1-2}
		$\mathcal{Q}_{T}$      & Sets of UAs allocated in the QoS class\\
		\cline{1-2}
		$\mathcal{G}_{Tt}$      & Sets of UAs allocated in the group t\\
		\cline{1-2}
		$\varsigma_{tm}$   & Sets of UAs allocated in mmW of the group t\\
		\cline{1-2}
		$\varsigma_{t\mu}$   & Sets of UAs allocated in $\mu$W of the group t\\
		\cline{1-2}
		$T$   & Number of slots of the QoS class $Q_T$\\
		\cline{1-2}
		$\mathcal{T}$   & Sets of slots of the QoS class $Q_T$\\
		\cline{1-2}
		$\mathcal{P}$   & Sets of minimum power over $\mu$W\\
		\cline{1-2}
		\hline
	\end{tabular}
\end{table}

\section{Group-based Algorithm}
All UAs from the same UE have the same transmission channel and suitable RBs. If these UAs are not grouped, they will compete for possession of the same RB in the same slot. It degrades system performance. To solve this problem, our proposed GB algorithm assigns these UAs to different time slots. Uncertainty, including $X_{\sigma}$ in large-scale pass loss, results from external factor are eliminated to ensure all UAs transmit in the same condition. Here, we take minimum power over $\mu$W as a metric. Besides, the number of UAs in each group becomes an important factor that affects transmission when BS needs to serve great quantity of UAs.
\subsection{Problem and Algorithm}
In this paper, the main goal of GB algorithm is to pursue the balance between overall power and the number of UAs for all time slots. Therefore, the problem can be formulated as:
\begin{subequations}
\begin{align}
&\mathop{\min}_{\mathcal{Q}_{T}} \sum_{t=1}^{T-1}\left(\left|\sigma_t\right|+\left|\epsilon_t\right|\right)\nonumber\\
&\text{ s.t.}\quad\frac{\left|\varsigma_t\right|}{\left|\varsigma_{(t+1)}\right|}+\sigma_t=1,(t={1,2...T-1}),\\
&\quad\quad\frac{\sum\limits_{n \in \mathcal{G}_{Tt}}\sum\limits_{k=1}^{K_2}{p_{1nkt}}}{\sum\limits_{n \in \mathcal{G}_{T(t+1)}}\sum\limits_{k=1}^{K_2}{p_{1nk(t+1)}}}+\epsilon_t=1,(t={1,2...T-1}),\\
&\quad\quad\sum\limits_{t=1}^{T} \left|\mathcal{G}_{Tt}\right| = N.
\end{align}
\end{subequations}

\begin{algorithm} 
	\caption{Group-based (GB) Algorithm} 
	\label{alg1} 
	\begin{algorithmic}[1] 
		\REQUIRE $\mathcal{P}$ in descending order\\
		number of groups $T$
		\ENSURE grouping situation
		\STATE Calculate the number of UAs expected in each group, $C_1$ and overall power, $C_2$;
		\FOR{$i=1$ to $N$}
		\IF {$i\leq T$}
		\STATE Add UA $i$ to group $i$;
		\ELSE
		\FOR{$j=1$ to $T$}
		\STATE Calculate deviation between current number of UAs in group $j$ and $C_1$, and the deviation is described as $a_1$;\\
		\STATE Calculate deviation between overall power of UAs in group $j$ and $C_2$, and the deviation is described as $a_2$;\\
		\STATE Assume that group $j$ gains UA $i$. Then deviation between current number of UAs in group $j$ and $C_1$ is calculated, and the deviation is described as $b_1$;\\
		\STATE Assume that group $j$ gains UA $i$. Then deviation between current overall power in group $j$ and $C_2$ is calculated, and the deviation is described as $b_2$;\\
		\STATE Record D($j$)=$\eta (a_1-b_1)+\gamma(a_2-b_2)$
		\ENDFOR
		\STATE Compare all elements in D and add UA $i$ to the most suitable group.
		\ENDIF
		\ENDFOR
	\end{algorithmic}
\end{algorithm}
\subsection{Parameter and Analysis}

In Algorithm 1, $\eta$ and $\gamma$ represent the degree of pursuit of the similar numbers and overall between groups, respectively. Algorithm 1 can adapt to various situations by adjusting the values of $\eta$ and $\gamma$. For instance, in the situation of small difference between $K_2$ and $N$, it is necessary to increase $\eta$ properly to ensure all UAs are served effectively.

\section{Allocation Algorithm in dual-mode}
To simplify the representations of $R_{nkt}^{(1)}$, $R_{nkt}^{(2)}$ ,$p_{1nkt}$ and $p_{2nkt}$, we define
\begin{equation*}
\begin{aligned}
N_{nkt}^{(1)}\triangleq\frac{\omega_{1}N_{0}}{\left|g_{kt}\right|^{2}10^{-0.1L_{1}(d_{n})}}, N_{nkt}^{(2)}\triangleq\frac{\omega_{2}N_{0}}{\psi(d_n)\left|h_{kt}\right|^{2}10^{-0.1L_{2}(d_{n})}},
\end{aligned}
\end{equation*}
then the rate over both bands can be expressed as:
\begin{equation*}
\begin{aligned}
R_{nkt}^{(1)}=\omega_{1}\log_{2}\left(1+\frac{p_{1nkt}}{N_{nkt}^{(1)}}\right),
R_{nkt}^{(2)}=\omega_{2}\log_{2}\left(1+\frac{p_{2nkt}}{N_{nkt}^{(2)}}\right),\\
p_{1nkt}=N_{nkt}^{(1)}\left(2^{\frac{R_{nkt}^{(1)}}{\omega_{1}}}-1\right),
p_{2nkt}=N_{nkt}^{(2)}\left(2^{\frac{R_{nkt}^{(2)}}{\omega_{2}}}-1\right).
\end{aligned}
\end{equation*}

Assume that UA $n$ initially occupy $E$ RBs over $\mu$W or $K_2$ RBs over mmW. To make sure (1c)-(1d) holds, we have 
\begin{equation*}
\begin{aligned}
\tau\sum\limits_{r=1}^{E}R_{nk_rt}^{(1)}=\tau\sum\limits_{r=1}^{E}\left[\omega_{1}\log_{2}(1+\frac{p_{1nk_rt}}{N_{nk_rt}^{(1)}})\right] \geq b_{nt}^{req},\\
\tau^{'}\sum\limits_{r=1}^{K_2}R_{nk_rt}^{(2)}=\tau^{'}\sum\limits_{r=1}^{K_2}\left[\omega_{2}\log_{2}(1+\frac{p_{2nk_rt}}{N_{nk_rt}^{(2)}})\right] \geq b_{nt}^{req}.
\end{aligned}
\end{equation*}

To satisfy the inequality of arithmetic and geometric means, we assume there are only $k_1$ RBs over $\mu$W or $k_2$ RBs over mmW available to transmit data for UA $n$. Therefore, the required minimum energy can be expressed as 
\begin{equation}
\begin{aligned}
p_{1nt}=\sum\limits_{r=1}^{K_1}p_{1nkt}=k_1\left({\prod\limits_{r=1}^{k_1}N_{nk_rt}^{(1)}2^{\frac{b_{nt}^{req}}{\tau^{'}\omega_{1}}}}\right)^\frac{1}{k_1}-\sum\limits_{r=1}^{k_1}N_{nk_rt}^{(1)},\\
p_{2nt}=\sum\limits_{r=1}^{K_2}p_{2nkt}=k_2\left({\prod\limits_{r=1}^{k_2}N_{nk_rt}^{(2)}2^{\frac{b_{nt}^{req}}{\tau^{'}\omega_{2}}}}\right)^\frac{1}{k_2}-\sum\limits_{r=1}^{k_2}N_{nk_rt}^{(2)}.
\end{aligned}
\end{equation}

Considering MILP with high complexity in each group,  decomposition the problem into dual frequency band can reduce the complexity. with several advantages of mmW, such as large bandwidth (for higher data transfer rates), low interference (systems with a high immunity to cramming), most UAs are inclined to be allocated over this band. UAs which are prone to be allocated to mmW have features of short-distance, less-obstacles, and low-power. At time slot $t$, mmW allocation is performed firstly and the corresponding problem can be extracted from (1):
\begin{subequations}
	\begin{align}
	&\mathop{\min}_{S_{nt},R_{nkt}^{(2)}} \sum_{n=1}^{N}\left(S_{nt}\sum_{k=1}^{K2}p_{2nkt}\right)\nonumber\\
	&\text{s.t.} \quad\sum\limits_{n=1}^{N} S_{nt} = N^{'},\\
	&\quad\quad\tau^{'}\sum\limits_{k=1}^{K_2}R_{nkt}^{(2)}=S_{nt}b_{nt}^{req},\forall n\in \mathcal{Q}_{T},\\
	&\quad\quad \mathcal{S} = \{S_{nt}|S_{nt}\in\{0,1\},\forall n\in \mathcal{Q}_{T}\},
	\end{align}
\end{subequations}
where $\sum_{k=1}^{K2}p_{2nkt}$ can be calculated by (3) directly. Now, there is only one variable left in the problem (4). Considering that UAs with low power are often more suitable for mmW, we use a greedy algorithm to solve the problem.

Next we schedule the rest of UAs in the group $t$ to $\mu$W and the problem is formulated as
\begin{subequations}
	\begin{align}
	&\mathop{\min}_{a_{nkt},R_{nkt}^{(1)}}\sum_{n=1}^{N}\left[\sum_{k=1}^{K1}\left(a_{nkt}p_{1nkt}\right)\right]\nonumber\\
	&\text{s.t.} \quad\sum\limits_{n=1}^{N} a_{nkt}\leq 1,\quad\forall n\in \varsigma_{t\mu},\forall k\in\mathcal{K}_1,\\
	&\quad\quad\tau\sum\limits_{k=1}^{K_1}{a_{nkt}R_{nkt}^{(1)}}=b_{nt}^{req},\forall n\in \varsigma_{t\mu},\\
	&\quad\quad \mathcal{A} = \{a_{nkt}|a_{nkt}\in\{0,1\},\forall n\in \varsigma_{t\mu},\forall k\in\mathcal{K}_1\},\\
	&\quad\quad R_{nkt}\geq 0,\quad\forall n\in \varsigma_{t\mu},\forall k\in\mathcal{K}_1.
	\end{align}
\end{subequations}

In the optimization problem (5), there exist two kinds of variables. $a_{nk}$ is 0-1 variable and $R_{nkt}^{(1)}$ is continuous variable. It turns out that the problem is a mixed integer nonlinear programming problem (MINLP). For further analysis, $a_{nk}$ determines whether UA $n$ obtains power in RB $k$. Here, the problem is simplified with the product relationship between $R_{nk}$ to replace (5a) (5c). Therefore, the optimization problem can be reformulated as:
\begin{subequations}
	\begin{align}
	&\mathop{\min}_{R_{nkt}^{(1)}} \sum_{n=1}^{N}\sum_{k=1}^{K1}p_{1nkt}\nonumber\\
	&\text{ s.t.}\quad\tau\sum\limits_{k=1}^{K_1}{R_{nkt}^{(1)}}=b_{nt}^{req},\quad\forall n\in \varsigma_{t\mu},\\
	&\quad\quad\quad\sum\limits_{i\neq j}R_{ikt}^{(1)}R_{jkt}^{(1)}=0,\quad\forall k\in\mathcal{K}_1,\\
	&\quad\quad\quad R_{nkt}^{(1)}\geq 0,\quad\forall n\in \varsigma_{t\mu},\forall k\in\mathcal{K}_1.
	\end{align}
\end{subequations}
By using Lagrange multipliers \cite{b10}, the Lagrangian is given by
\begin{equation*}
\begin{aligned}
L&=\sum\limits_{n=1}^{N}\sum\limits_{k=1}^{K_1}p_{1nkt}+\sum\limits_{n=1}^{N}\beta_{n}\left(\tau\sum\limits_{k=1}^{K_1}R_{nkt}^{(1)}-b_{nt}^{req}\right)\\
&+\sum\limits_{k=1}^{K_1}\lambda_k\left(\sum\limits_{i\neq j}R_{ikt}^{(1)}R_{jkt}^{(1)}\right)+\sum\limits_{n=1}^{N}\sum\limits_{k=1}^{K_1}\mu_{nk}\left(-R_{nkt}^{(1)}\right),
\end{aligned}
\end{equation*}
where $\beta_{n}$, $\lambda_k$ and $\mu_{nk}$ are the Lagrangian multipliers for the constraints (6a)-(6c), respectively. $\mu_{nk}$ must be non-negative.

After differentiating $L$ with respect to $R_{1nkt}$, the necessary condition for optimal solution, $R_{nkt}^{(1)*}$ and $R_{ikt}^{(1)*}$, is shown as follows:
\begin{equation*}
\begin{aligned}
\frac{\partial L}{\partial R_{nkt}^{(1)}}&=\frac{N_{nkt}^{(1)}}{\omega_1}2^{\frac{R_{nkt}^{(1)*}}{\omega_1}}\ln{2}+\beta_{n}\tau+\lambda_k\left(\sum\limits_{i\neq n}R_{ikt}^{(1)*}\right)-\mu_{nk}\\
&=0.
\end{aligned}
\end{equation*}

Specially, if $R_{nkt}^{(1)*} \neq 0$, we can get $R_{ikt}^{(1)*}=0$ for all $i \neq n$ from (6b). The relationship between $R_{nkt}^{(1)*}$ and $\beta_{n}$ is given by
\begin{equation}
\begin{aligned}
R_{nkt}^{(1)*} \geq \omega_{1} \log_{2}{\frac{-\beta_{n}\tau\omega_{1}}{N_{nkt}^{(1)}\ln 2}}.
\end{aligned}
\end{equation}

We assume that $m$ RBs are allocated to UA $n$. To guarantee UA $n$ satisfies (6a), we have
\begin{equation}
\begin{aligned}
\log_{2}{(-\beta_{n})}\geq\frac{b_{nt}^{req}}{\tau m \omega_{1}}-\frac{\sum\limits_{k=1}^{m}\log_{2}\frac{\tau\omega_{1}}{N_{nkt}^{(1)}\ln 2}}{m}.
\end{aligned}
\end{equation}

Unfortunately, $m$ is so hard to be solved that we cannot get the correct value of $\log_{2}(-\beta_{n})$. Qualitatively, it is possible for long-distance UAs to require more RBs, owing to high $N_{nkt}^{(1)}$. Therefore, the initial $m$ is proportional to distance. The first part is replaced by $\frac{b_{nt}^{req}}{\tau m_n \omega_{1}}$ as an initial value, where $m_n$ represents the number of RBs allocated to UA $n$; For the second part, we take mean value of all RBs, $\frac{\sum\limits_{k=1}^{K_1}\log_{2}\frac{\tau\omega_{1}}{N_{nkt}^{(1)}\ln 2}}{K_1}$, to replace $\frac{\sum\limits_{k=1}^{m}\log_{2}\frac{\tau\omega_{1}}{N_{nkt}^{(1)}\ln 2}}{m}$.


Therefore, $\log_{2}(-\beta_{n})$ is initialized as:
\begin{equation}
\log_{2}(-\beta_{n})_{\rm init}=\frac{b_{nt}^{req}}{\tau m_n \omega_{1}}-\frac{\sum\limits_{k=1}^{K_1}\log_{2}\frac{\tau\omega_{1}}{N_{nkt}^{(1)}\ln 2}}{K_1}.
\end{equation}

Then, we substitute (9) into (7) and restrict $R_{nkt}^{(1)}$ to be greater than zero. Once $R_{nkt}^{(1)}$ is determined, the problem in (5) becomes a 0-1 integer linear programming (0-1 ILP) problem, which  can be formulated as:
\begin{subequations}
\begin{align}
&\mathop{\min}_{a_{nkt}}\sum_{n=1}^{N}\left[\sum_{k=1}^{K1}\left(a_{nkt}p_{1nkt}\right)\right]\nonumber\\
&\text{s.t.} \quad\sum\limits_{n=1}^{N} a_{nkt}\leq 1,\quad\forall n\in \varsigma_{t\mu},\forall k\in\mathcal{K}_1,\\
&\quad\quad\tau\sum\limits_{k=1}^{K_1}{a_{nkt}R_{nkt}^{(1)}}\geq b_{nt}^{req},\forall n\in \varsigma_{t\mu},\\
&\quad\quad \mathcal{A} = \{a_{nkt}|a_{nkt}\in\{0,1\},\forall n\in \varsigma_{t\mu},\forall k\in\mathcal{K}_1\}.
\end{align}
\end{subequations}
Increasing $\log_{2}(-\beta_{n})$ with the step of $\Delta$ until feasible region of (10) is nonempty, we have 
\begin{equation}
\log_{2}(-\beta_{n})=\log_{2}(-\beta_{n})+\Delta.
\end{equation}

Suppose that UA $n_1$ and UA $n_2$ own sets of RBs  $\mathcal{I}_{m_1}=\{k_{I_1},k_{I_2},\cdots,k_{I_{m_1}}\}$ and $\mathcal{J}_{m_2}=\{k_{J_1},k_{J_2},\cdots,k_{J_{m_2}}\}$, respectively. Each RB cannot be allocated to different UAs, i.e., $\mathcal{I}_{m_1} \cap \mathcal{J}_{m_2}= \emptyset$.
According to (3), the minimum energy of both UAs over $\mu$W is given by
\begin{equation*}
\begin{aligned}
V(n_1,\mathcal{I}_{m_1})=m_1\left({\prod\limits_{r=1}^{m_1}N_{n_1k_{I_r}t}^{(1)}2^{\frac{b_{n_1t}^{req}}{\tau\omega_{1}}}}\right)^\frac{1}{m_1}-\sum\limits_{r=1}^{m_1}N_{n_1k_{I_r}t}^{(1)},\\
V(n_2,\mathcal{J}_{m_2})=m_2\left({\prod\limits_{r=1}^{m_2}N_{n_2k_{J_r}t}^{(1)}2^{\frac{b_{n_2t}^{req}}{\tau\omega_{1}}}}\right)^\frac{1}{m_2}-\sum\limits_{r=1}^{m_2}N_{n_2k_{J_r}t}^{(1)}.\\
\end{aligned}
\end{equation*}

Now, the ownership of RB $k_{I_{m_1}}$ transfers from UA $n_1$ to $n_2$. UA $n_1$ owns sets of RBs $\mathcal{I}_{m_1}^{'}=\{k_{I_1},k_{I_2},\cdots,k_{I_{m_1-1}}\}$. Then, the variation of power is given by
\begin{equation}
\begin{aligned}
&V(n_1,\mathcal{I}_{m_1})-V(n_1,\mathcal{I}_{m_1}^{'})=m_1\left({\prod\limits_{r=1}^{m_1}N_{n_1k_{I_r}t}^{(1)}2^{\frac{b_{n_1t}^{req}}{\tau\omega_{1}}}}\right)^\frac{1}{m_1}-\\
&(m_1-1)\left({\prod\limits_{r=1}^{m_1-1}N_{n_1k_{I_r}t}^{(1)}2^{\frac{b_{n_1t}^{req}}{\tau\omega_{1}}}}\right)^\frac{1}{m_1-1}-N_{n_1k_{I_{m_1}}t}^{(1)}.
\end{aligned}
\end{equation}

Similarly, UA $n_2$ owns sets of RBs $\mathcal{J}_{m_2}^{'}=\{k_{J_1},k_{J_2},\cdots,k_{J_{m_2}},k_{I_{m_1}}\}$. Normally, the inequality of arithmetic and geometric means holds and the variation of power is given by
\begin{equation}
\begin{aligned}
&V(n_2,\mathcal{J}_{m_2})-V(n_2,\mathcal{J}_{m_2}^{'})=m_2\left({\prod\limits_{r=1}^{m_2}N_{n_2k_{J_r}t}^{(1)}2^{\frac{b_{n_2t}^{req}}{\tau\omega_{1}}}}\right)^\frac{1}{m_2}-\\
&(m_2+1)\left({\prod\limits_{r=1}^{m_2}N_{n_2k_{J_r}t}^{(1)}N_{n_2k_{I_{m_1}}t}^{(1)}2^{\frac{b_{n_2t}^{req}}{\tau\omega_{1}}}}\right)^\frac{1}{m_2+1}+N_{n_2k_{I_{m_2}}t}^{(1)},
\end{aligned}
\end{equation}
Otherwise, we have
\begin{equation}
\begin{aligned}
V(n_2,\mathcal{J}_{m_2})-V(n_2,\mathcal{J}_{m_2}^{'})=0.
\end{aligned}
\end{equation}

According to (12)-(14), it is easy to obtain the matrix ${\bf A}$  recording gain and loss at each RB. So, any change of power at each RB, resulting from ownership transferring, can be obtained by adding elements in {\bf A}. And the largest reduction at each RB is stored in matrix {\bf B}. At every iterate, we choose which RB's ownership should be transferred according to {\bf B}.

\begin{algorithm} 
	\caption{Allocation Scheme in Dual-mode Base Station, EOD algorithm} 
	\label{alg2} 
	\begin{algorithmic}[1] 
		\REQUIRE $N_{nkt}^{(1)}$, $N_{nkt}^{(2)}$, $b_{nt}^{req}$, $N^{'}$, $\Delta$
		\ENSURE Allocation in mmW and $\mu$W;
		\STATE Calculate power of UAs need in mmW with (3) and sort them in ascending order;
		\FOR{$i=1$ to $N^{'}$}
		\STATE Assign UA $i$ to mmW;
		\ENDFOR
		\STATE Remove UAs already allocated in mmW and the rest of UAs are allocated to $\mu$W;
		\STATE Initialize $\log_{2}(-\beta_{n})$ with (9) for all UAs;
		
		\WHILE{Feasible region in (10) is empty}
		\STATE Increase $\log_{2}(-\beta_{n})$ in the step of $\Delta$;
		\ENDWHILE
		
		\STATE Adjust power to reach minimization with (3);
		\STATE Calculate ${\bf A}$ and ${\bf B}$ by gaining and losing at all RBs;
		\STATE Transfer ownership the maximum element in ${\bf B}$;
		\REPEAT 
		\STATE{Step 10-12}
		\UNTIL {Any RB's ownership transferred leads to power increasing. }
	\end{algorithmic}
\end{algorithm}

\section{Simulation Results}
In this section, we evaluate the performance of our proposed GB-EOD algorithm scheme in a multiuser downlink dual-mode transmission system by comparing it with context-aware algorithm. As in \cite{b3}, we assume BS is located at the center of the small cell as (0, 0). Besides, $M$ UEs are assumed to be uniformly distributed within 5-200 meters away from BS. Each UE runs $\kappa$ UAs independently. Some simulation parameters are listed in TABLE II.
  
\begin{table}
\centering
\caption{Some simulation parameters}
\begin{tabular}{ |c|c| }
	\hline  
	Parameters & Value \\
	\cline{1-2}
	Transmit bits for each UA, $b_n^{req}$ & 10Kbits\\
	\cline{1-2}
	Available bandwidth, $\Omega_{1},\Omega_{2}$ & 10MHz,1GHz \\
	\cline{1-2}
	Bandwidth per RB, $\omega_{1},\omega_{2}$ & 180KHz,180KHz \\
	\cline{1-2}
	Rician K-factor & 2.4 \\
	\cline{1-2}
	large-scale channel effects, $\alpha_{1},\alpha_{2}$ & 38dB,70dB \\
	\cline{1-2}
	large-scale channel effects, $\beta_{1},\beta_{2}$ & 3,2 \\
	\cline{1-2}
	large-scale channel effects, $X_{\sigma_{1}},X_{\sigma_{2}}$ & 10,5.2 \\
	\cline{1-2}
	Antenna gain, $\psi$ & 18dBi \\
	\cline{1-2}
	Time slot duration, $\tau$ & 10ms \\
	\cline{1-2}
	Beam-training overhead, $\tau^{'}$ & 0.1ms \\
	\cline{1-2}
	Number of UAs per UE, $\kappa$ & 3 \\
	\cline{1-2}
	Number of UAs in mmW at each slot, $N^{'}$ & 20 \\
	\cline{1-2}
	\cline{1-2}
	Increasing step, $\Delta$ & 0.01 \\
	\cline{1-2}
	\hline
\end{tabular}
\end{table}

\subsection{Power vs Number of UEs}
\begin{figure}[htbp]
	\centerline{\includegraphics[scale=0.65]{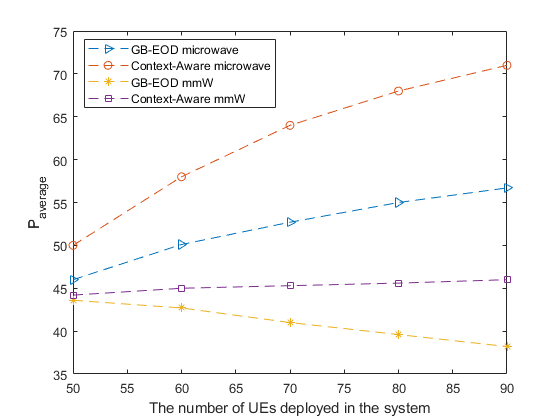}}
	\caption{Power comparison for two algorithm over both frequency bands}
	\label{fig:3}
\end{figure}

Fig.~\ref{fig:3} compares the performance of context-aware algorithm and GB-EOD algorithm. Under the premise of fixed power allocated to mmW in both algorithm, we compare the power required over $\mu$W band. It shows that low power is required in GB-EOD algorithm. It is reasonable because GB-EOD scheme in the same power can serve more UAs with proper power allocation, thus reduce the transmitting pressure on the other band. It is similar with fixed power over $\mu$W.

\subsection{Power vs Number of UAs allocated over mmW}
\begin{figure}[htbp]
	\centerline{\includegraphics[scale=0.65]{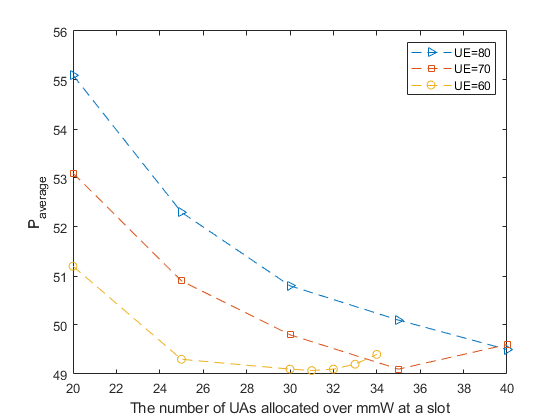}}
	\caption{UAs allocated in mmW increasing from 20 to 40}
	\label{fig:4}
\end{figure}

Curves in Fig.~\ref{fig:4} trace out effect of $N^{'}$. It shows how changes in $N^{'}$ will affect the power consumption. As the number of UAs increases from 20 to 40, the average power first decrease and then increase. Downward power consumption curve results from more spectrum and beamforming technology over mmW band. Oppositely, upward power consumption curve implies the crowding over mmW. Therefore, we have reasons to believe that there exit a extreme point in each curve. The more UEs is, the smaller extreme point is. 

\section{Conclusion}
In this paper, multiuser downlink dual-mode system has been studied to minimize the power. Before allocated in dual-mode, UAs are divided into multiple groups for the propose of decreasing competition. In each group, all UAs must transmit successfully to ensure fairness. Over mmW band, we take a greedy algorithm to select short-distance and low-power UAs to perform efficiently. Over $\mu$W band, we estimate $\beta_{n}$ by KKT conditions and take low-complexity algorithm to reduce power. Considering the performance of system, GB-EOD algorithm implements a pretty tradeoff between fairness and low power.

\end{document}